\documentclass[letterpaper,12pt]{article}

\usepackage[dvips]{graphicx}
\usepackage[pdftex]{color}
\usepackage{ifthen}
\usepackage{enumerate}
\usepackage{cite}
\usepackage{amssymb}
\usepackage[mathscr]{eucal}
\usepackage[errorshow]{tracefnt}
\usepackage{amsmath}
\usepackage{amssymb}
\usepackage{amsthm}
\usepackage{eucal}
\usepackage{eufrak}

\newcommand{\dlb}{\ensuremath{[\![}}
\newcommand{\drb}{\ensuremath{]\!]}}

\newcommand{\cA}{{\cal A}}

\newcommand{\cG}{{\cal G}}

\newcommand{\cL}{{\cal L}}

\newcommand{\cN}{{\cal N}}

\newcommand{\cZ}{{\cal Z}}

\newcommand{\beq}{\begin{equation}}
\newcommand{\beqn}{\begin{equation*}}
\newcommand{\eeq}{\end{equation}}
\newcommand{\eeqn}{\end{equation*}}
\newcommand{\beqa}{\begin{eqnarray}}
\newcommand{\beqan}{\begin{eqnarray*}}
\newcommand{\eeqa}{\end{eqnarray}}
\newcommand{\eeqan}{\end{eqnarray*}}
\newcommand{\bdm}{\begin{displaymath}}
\newcommand{\edm}{\end{displaymath}}

\newcommand{\la}{\langle}
\newcommand{\ra}{\rangle}

\newcommand{\ba}{\begin{array}}
\newcommand{\ea}{\end{array}}

\newcommand\ffam{\sffamily}
\newcommand\fser{\bfseries}

\newcommand\nn{\nonumber}

\newcommand\benu{\begin{enumerate}}
\newcommand\eenu{\end{enumerate}}
\newcommand\bit{\begin{itemize}}
\newcommand\eit{\end{itemize}}

\def\der'{\mathfrak{der}'\,}
\def\der{\mathfrak{der}\,}
\def\str'{\mathfrak{str}'\,}
\def\str{\mathfrak{str}\,}

\newcommand{\al}{\alpha}
\newcommand{\be}{\beta}
\newcommand{\ga}{\gamma}

\newcommand{\de}{\delta}

\newcommand{\om}{\omega}
\newcommand{\Om}{\Omega}

\def\xone{I_{ABC}}
\def\xtwo{I_{BCA}}
\def\xthree{I_{CAB}}
\def\xfour{I_{BAC}}
\def\xfive{I_{CBA}}
\def\xsix{I_{ACB}}

\def\xeight{I_{LC}}
\def\xnine{I_{LR}}
\def\xninehalf{I_{RL}}
\def\xten{I_{CL}}

\def\xtwelve{I_{CR}}
\def\xtwelvehalf{I_{RC}}
\def\xthirteen{I_{RR}}

\RequirePackage{hyperref}

\numberwithin{equation}{section}

\begin{document}

\hfill{\tt ULB-TH/11-03}\\
\vskip-10pt

\pagestyle{empty}

\begin{center}

\vspace*{1cm}

\noindent
\begin{center}
{\LARGE \textsf{\textbf{Unifying $\cN=5$ and $\cN=6$}} }\\
\vspace{.2cm}\hspace{.1cm}

\vskip 1truecm

{\large \textsf{\textbf{Jakob Palmkvist}}} \\
\vskip 1truecm
        {\ffam
        {Physique Th\'eorique et Math\'ematique\\
  Universit\'e Libre de Bruxelles \& International Solvay Institutes\\
  Boulevard du Triomphe, Campus Plaine, ULB-CP 231,\\BE-1050 Bruxelles, Belgium}\\[3mm]}
  \textsf{and}\\[3mm]
        {\ffam
        {Department of Fundamental Physics\\
  Chalmers University of Technology,\\SE-412 96 G\"oteborg, Sweden}\\[3mm]}
        {\tt jakob.palmkvist@ulb.ac.be} \\

\end{center}

\vskip 1cm

\centerline{\ffam\fser Abstract}
\end{center}
We write the Lagrangian of the general $\cN=5$ three-dimensional superconformal Chern-Simons theory, based on a basic Lie superalgebra,
in terms of our recently introduced 
$\cN=5$ three-algebras. These include $\cN=6$ and $\cN=8$ three-algebras as special cases. When we impose an antisymmetry condition on the triple product, the supersymmetry automatically enhances, and the $\cN=5$ Lagrangian reduces to that of the well known 
$\cN=6$ theory, including the ABJM and ABJ models.

\newpage

\pagestyle{plain}

\section{Introduction}

Following the construction by Bagger, Lambert \cite{Bagger:2006sk,Bagger:2007jr,Bagger:2007vi} and Gustavsson \cite{Gustavsson:2007vu} of a three-dimensional superconformal Chern-Simons theory with maximal ($\cN=8$) supersymmetry, similar theories with less supersymmetry have subsequently been constructed \cite{Aharony:2008ug,Benna:2008zy,Hosomichi:2008jb,Bagger:2008se,Bergshoeff:2008bh,Aharony:2008gk,deMedeiros:2009eq,Chen:2009cwa,Bagger:2010zq,Gustavsson:2010yr} (including the ABJM \cite{Aharony:2008ug} and ABJ \cite{Aharony:2008gk} models).

The interest in three-dimensional superconformal theories has been motivated by the search for a world-volume theory that could describe interacting M2-branes, and hopefully give a better understanding of M-theory. The concept of triple products has played an important role in that search, and we believe that it will be useful also in the future.

In this paper we focus on the Lagrangian descriptions of the (classical) $\cN=6$ and $\cN=5$ theories given in
\cite{Bagger:2008se,Chen:2009cwa} in terms of anti-Lie triple systems.
We will show that if one instead uses the 
$\cN=5$ three-algebras introduced in \cite{Kim:2010kq},
then the $\cN=5$ theory and the $\cN=6$ theory can be written in exactly the same way.
When we impose an antisymmetry condition on the triple product, reducing the $\cN=5$ three-algebra to an
$\cN=6$ three-algebra, the supersymmetry automatically enhances, and the
$\cN=5$ theory reduces to the $\cN=6$ theory.

Before proceeding, a few words about our terminology are in order.
In this paper we use the term {\it triple system} for any vector space with a trilinear {triple product}.
By equipping the triple system with an inner product, and imposing certain conditions on the 
triple product and the inner product, we obtain special cases that we call {\it three-algebras}.
Among these, the $\cN=5$ three-algebras are the most general. They include 
$\cN=6$ three-algebras, which in turn include $\cN=8$ three-algebras. The $\cN=8$ and $\cN=6$ three-algebras are the ones originally introduced by Bagger and Lambert in their constructions of $\cN=8$ \cite{Bagger:2006sk,Bagger:2007jr,Bagger:2007vi}
and $\cN=6$ \cite{Bagger:2008se} theories. A similar construction of an $\cN=5$ theory was given in \cite{Chen:2009cwa,
Bagger:2010zq}, based on a triple system that was called a ({symplectic}) `$\cN=5$ three-algebra'.
However, this does {not} agree with the definition of an $\cN=5$ three-algebra in 
\cite{Kim:2010kq}, which we follow in this paper (up to a slight modification, see below).

Unlike the $\cN=5$ three-algebras in 
\cite{Kim:2010kq}, the triple systems considered in \cite{Chen:2009cwa,Bagger:2010zq,2010arXiv1010.3599C} do not include the $\cN=6$ and $\cN=8$ three-algebras as special cases.
They are triple systems of a different kind, called 
{\it anti-Lie triple systems} and well known by their connection to Lie superalgebras.
It was first noted in  
\cite{deMedeiros:2008zh} that 
anti-Lie triple systems are relevant for $\cN=5$ theories, and this idea was further elaborated in 
\cite{FigueroaO'Farrill:2009pa,deMedeiros:2009eq}.

We will see in this paper that there is no need of introducing a different kind of triple system, one can just generalize the 
$\cN=6$ three-algebra to an $\cN=5$ three-algebra.
Our derivation of the $\cN=6$ theory from the $\cN=5$ theory is analogous to the one in \cite{Chen:2009cwa}, but instead of rewriting the expressions we just keep them as they are, and impose antisymmetry of the triple product.
Thus the embedding of the $\cN=6$ theory becomes manifest.

The paper is organized as follows. 
In section 2 we define the triple systems that we use, three-algebras and anti-Lie triple systems, and explain their connection to Lie superalgebras. In section 3 we present our result, the $\cN=5$ three-algebra formulation of the $\cN=5$ Lagrangian. We then show in section 4 how this result can be obtained from the formulation in \cite{Chen:2009cwa}, based on anti-Lie triple systems. In section 5 we impose the antisymmetry condition, and show that our result then automatically becomes the well known 
$\cN=6$ three-algebra formulation of the $\cN=6$ Lagrangian. We end with a conclusion in section 6.

\section{Triple systems} 

For a comprehensive review of triple systems (and other $n$-ary algebras), see \cite{2010JPhA...43C3001D}.

\subsection{Three-algebras}
\label{three-alg-subsection}
An $\mathcal{N}=5$ { three-algebra} is a triple system $V$ with a trilinear triple product
\begin{align}
V \times V \times V &\to V, & (x,y,z) \mapsto (xyz)
\end{align}
and an {inner product}, a symmetric bilinear form
\begin{align}
V \times V &\to \mathbb{C}, & (x,y) \mapsto \la x | y \ra,
\end{align}
that satisfies the {\it fundamental identity}
\begin{align}
(uv(xyz))-(xy(uvz))&=((uvx)yz)-(x(vuy)z),\label{N=5ta1}
\end{align}
the `generalized antisymmetry condition'
\begin{align}
K_{xy}(K_{uv}(z))&=(K_{xy}(v)uz)+(K_{xy}(u)vz),
\label{N=5ta2}
\end{align}
where $K_{xy}(z)=(xzy)+(yzx)$,
and invariance of the inner product,
\begin{align}
\la w | (xyz)\ra=\la y | (zwx) \ra&=\la(wzy)|x\ra=\la(yxw)|z\ra. \label{N=5ta3}
\end{align}

By imposing further conditions one obtains $\cN=6$ and $\cN=8$ three-algebras.
An $\mathcal{N}=6$ { three-algebra} is an $\mathcal{N}=5$ three-algebra with $K_{xy}=0$ for any pair $(x,\,y)$. This means that the triple product is antisymmetric under permutation of the first and the third argument,
\begin{align}
(xzy)=-(yzx), \label{antisymmetrycond}
\end{align}
so that (\ref{N=5ta2}) is trivially satisfied in an $\mathcal{N}=6$ { three-algebra}.
An $\mathcal{N}=8$ { three-algebra} is an $\mathcal{N}=6$ three-algebra where the triple product is 
also antisymmetric under permutation of the first and the second argument, and thus
totally 
antisymmetric.

It is worth mentioning that if we remove the minus sign in (\ref{antisymmetrycond}) and impose this {\it symmetry} condition on the triple product, then we obtain a {\it Jordan triple system} instead. Replacing it with a generalization corresponding to 
(\ref{N=5ta2}) gives a {\it Kantor triple system}, and removing it gives a {\it generalized Jordan triple system}. It was first noted in \cite{Nilsson:2008kq} that an $\cN=6$ three-algebra in fact is a generalized Jordan triple system.

The difference in the definitions here compared to those in \cite{Palmkvist:2009qq,Kim:2010kq} is that the triple product and inner product in an $\cN=5$ or $\cN=6$ three-algebra are now supposed to be linear in each of the arguments, instead of antilinear in some of them, and the inner product is supposed to be symmetric. This simpler setting is enough, at least for the purposes in this paper. If needed, anti-linearity can always be restored by the insertion of a conjugation in the triple product and the inner product, as explained in \cite{Palmkvist:2009qq}.
A difference compared to \cite{Bagger:2008se,Chen:2009cwa,Bagger:2010zq} is 
the order of the arguments in the triple product, which follows 
\cite{Nilsson:2008kq,Palmkvist:2009qq,Kim:2010kq}.
We will see the reason for this soon.

\subsection{Anti-Lie triple systems} 
An anti-Lie triple system
is a triple system $V$ with a trilinear triple product 
\begin{align}
V \times V \times V &\to V, & (x,y,z) \mapsto [xyz]
\end{align}
that
satisfies another `fundamental identity',
\begin{align}
[uv[xyz]]-[xy[uvz]]=[[uvx]yz]+[x[uvy]z],\label{alts1}
\end{align}
the symmetry condition
\begin{align}
[xyz]=[yxz] \label{alts2}
\end{align}
and the {\it cyclicity} condition
\begin{align}
[xyz]+[yzx]+[zxy]&=0.\label{alts3}
\end{align}
The anti-Lie triple systems in \cite{FigueroaO'Farrill:2009pa,Chen:2009cwa,Bagger:2010zq} 
furthermore have an invariant antisymmetric bilinear form, and the {\it quaternionic} anti-Lie triple systems in 
\cite{FigueroaO'Farrill:2009pa} also have a quaternionic structure.

\subsection{Connection to 5-graded Lie superalgebras}
\label{Connection to 5-graded Lie superalgebras}
The odd subspace of any Lie superalgebra is an anti-Lie triple system with the triple product
\begin{align}
[xyz]=[\{x,\,y\},z],
\end{align}
and conversely, any anti-Lie triple system gives rise to an associated Lie superalgebra \cite{2010arXiv1010.3599C}.
We consider here anti-Lie triple systems such that
the associated Lie superalgebra is a {\it basic} Lie superalgebra, that is, one of 
$A(m,n)$, $B(m,n)$, $C(n+1)$, $D(m,n)$, $F(4)$, $G(3)$, $D(2,1;\alpha)$ in the classification by Kac \cite{Kac77A}.
These are the ones that appear in \cite{Bagger:2010zq}, and in \cite{FigueroaO'Farrill:2009pa} they were shown to be in one-to-one correspondence with simple positive-definite quaternionic anti-Lie triple systems. As noted in \cite{2010arXiv1010.3599C} there is an additional class of simple Lie superalgebras with a non-degenerate bilinear form, denoted $H(2k)$, for $k\geq3$. Although the corresponding anti-Lie triple system satisfies the requirements in \cite{Chen:2009cwa,Bagger:2010zq}, it seems difficult to make sense of a theory based on such a triple system. The even subalgebra of $H(2k)$ (which would be the gauge algebra) is not semi-simple, and the adjoint representation on the odd subspace (in which the matter fields would transform) is not fully reducible.

In \cite{Palmkvist:2009qq} (type I) and \cite{Kim:2010kq} (type II) we explicitly studied each of the basic Lie superalgebras in the context of $\cN=6$ and $\cN=5$ three-algebras. Here we keep the discussion very general but it is intended to be applied to the special cases considered in \cite{Palmkvist:2009qq,Kim:2010kq}.

Let $\cG$ be a basic Lie superalgebra. Then it has a 5-grading, which 
means that it can be written as a direct sum of subspaces $\cG_k$ for all integers 
$k$ such that $\dlb \cG_i, \cG_j \drb\subset \cG_{i+j}$, and $\cG_k=0$ for $|k|\geq 3$. We will also allow for the possibility that $\cG_{-2}=\cG_{2}=0$, which means that we consider 3-graded Lie superalgebras as special cases of 5-graded ones. The even and odd parts consist of those subspaces $\cG_k=0$ for which $k$ is even and odd, respectively,
\begin{align} \label{evenoddparts}
\cG_{(0)}&=\cG_{-2}+\cG_{0}+\cG_{2}, & \cG_{(1)}&=\cG_{-1}+\cG_{1}.
\end{align}

In a basic Lie superalgebra we can furthermore define
a non-degenerate inner product $\Omega$ (a consistent and supersymmetric invariant bilinear form \cite{Frappat}) 
such that $\Om(\cG_i,\cG_j)=0$ unless $i+j=0$.
We will henceforth write $\Omega(x,y)=(x|y)$.
Let $T_a$ be a basis of $\cG_{(1)}$ and split the indices $a$ into $\alpha,\dot\al$, so that
$T_\al$ and $T_{\dot\al}$ are bases of $\cG_{1}$ and $\cG_{-1}$ respectively.
Then $\Om_{ab}=\Om(T_a,T_b)$ is zero unless one index is dotted an one undotted when we decompose the latin indices into greek ones. Note that $\Om_{ab}$ is antisymmetric, $\Om_{ab}=-\Om_{ba}$.

We use $\Om_{ab}$ and its inverse $\Om^{ab}$ (defined by $\Om^{ac}\Om_{cb}=\de^a{}_b$) to raise and lower indices from the left. Thus we can replace dotted indices downstairs with undotted indices upstairs, and write the basis of $\cG_{-1}$ as $T^\al$.

The inner product $\Omega$ induces an automorphism $\tau$ on $\cG$ given on $\cG_{(1)}$ by
\begin{align}
\tau(T^\al)&=T_\al,&\tau(T_\al)&=-T^\al.
\end{align}
It is then given by $\cG_{(0)}$ by the homomorphism property 
$\tau(\dlb x,y \drb)=\dlb \tau(x),\tau(y)\drb$. This automorphism satisfies
$\tau^2(x)=(-1)^px$ if $x \in \cG_{(p)}$, like the {\it graded superconjugation} in \cite{Palmkvist:2009qq}.
However, in accordance with our new convention for the triple product we let it be linear instead of antilinear.

Using the automorphism $\tau$ we can define a triple product on $\cG_{-1}$ by
\begin{align} \label{fromlsaton5}
(xyz)=[\{x,\tau(y)\},z],
\end{align}
and it is straightforward to check that $\cG_{-1}$ is an $\cN=5$ three-algebra with this triple product, and with the inner product
\begin{align}
\la x|y\ra=-(x|\tau(y)).
\end{align}
Furthermore, it follows from the Jacobi superidentity that this triple product is antisymmetric in the first and third argument (so that the $\cN=5$ three-algebra 
becomes an $\cN=6$ three-algebra) if and only if $\cG_{\pm2}=0$.

When we reformulate the $\cN=5$ theory we start with the anti-Lie triple system used in the original formulation, go to the associated Lie superalgebra (which we assume to be basic), and then from the Lie superalgebra to the $\cN=5$ three-algebra by (\ref{fromlsaton5}). To this end, we introduce structure constants for the anti-Lie triple system,
\begin{align}
f_{abcd}&=-([T_a T_b T_c]|T_d)=-([\{T_a, T_b\}, T_c]|T_d),
\end{align}
\begin{align}
[T_a T_b T_c]&=[\{T_a, T_b\}, T_c]=f_{abc}{}^d T_d
\end{align}
and write the identities (\ref{alts1})--(\ref{alts3}) in component form,
\begin{align}
f_{abf}{}^gf_{cde}{}^f-f_{cdf}{}^gf_{abe}{}^f&=
f_{abc}{}^f f_{fde}{}^g+f_{abd}{}^f f_{cfe}{}^g, \label{altscomp1}
\end{align}
\begin{align}
f_{abcd}&=f_{bacd},&
f_{abcd}+f_{bcad}+f_{cabd}&=0. \label{altscomp2}
\end{align}
In addition, we have the identity
\begin{align} 
f_{abcd}=f_{cdab} \label{altscomp3}
\end{align}
from the invariance of the inner product in the Lie superalgebra.

It follows from the 5-grading that when we split the latin indices into greek ones only the structure constants with two dotted and two undotted indices survive. Furthermore, we can fix the position of the two dotted indices to the first and the third by using (\ref{altscomp2})--(\ref{altscomp3}),
\begin{align}
f_{\dot\al\be\de\dot\ga}=f_{\be\dot\al\dot\ga\de}&=f_{\be\dot\al\de\dot\ga}=f_{\dot\al\be\dot\ga\de},\nn\\
f_{\ga\de\dot\al\dot\be}=f_{\dot\al\dot\be\ga\de}&=-f_{\dot\al\ga\dot\be\de}-f_{\dot\be\ga\dot\al\de}.
\label{altscomp4}
\end{align}
Finally, we can raise the dotted indices using $\Omega$ and remove the dots,
\begin{align}
f^\al{}_\be{}^\ga{}_\de = \Om^{\al\dot\al}\Om^{\ga\dot\ga}f_{\dot\al\be\dot\ga\de}.
\end{align}
Then we end up with the structure constants of the $\cN=5$ three-algebra $\cG_{-1}$, 
defined by (\ref{fromlsaton5}). Indeed,
\begin{align}
f^\al{}_\be{}^\ga{}_\de &= -\big([T^\al T_\be T^\ga]\big|T_\de\big)=-\big([\{T^\al,T_\be\},T^\ga]\big|T_\de\big)\nn\\
&=-\big([\{T^\al,\tau(T^\be)\},T^\ga]\big|\tau(T^\de)\big)=-\big\la(T^\al T^\be T^\ga)\big|T^\de\big\ra,
\end{align}
\begin{align}
\la T^\al | T^\be \ra &= -(T^\al| \tau(T^\be))=-(T^\al|T_\be)\nn\\&=-\Om_{\be\dot\be}(T^\al|T^{\dot\be})
=-\Om_{\be\dot\be}\Om^{\al\dot\be}=\de^\al{}_\be.
\end{align}
This explains why
we write the structure constants for a three-algebra
with indices alternating upstairs and downstairs, and in a different order compared to 
\cite{Bagger:2008se,Chen:2009cwa,Bagger:2010zq}.
It was pointed out already in \cite{Nilsson:2008kq} that it is natural to place the indices in this way.

\section{The $\cN=5$ theory}

We denote the matter fields by $\cZ^A{}_a, \Psi_A{}^a$ in the $\cN=5$ theory, and by
$Z^A{}_\alpha, \psi_A{}^\alpha$ in the $\cN=6$ theory. The indices $A,B,\ldots$ refer to the 
R-symmetry, which is $\rm{Sp}(4)$ in the $\cN=5$ theory and $\rm{SU}(4)$ in the $\cN=6$ theory.
In both cases they take the values $1,2,3,4$.
We will always raise and lower the R-symmetry indices with
the $\rm{Sp}(4)$ invariant tensor $\omega$ and its inverse, for example
\begin{align}
\cZ_A{}_a &= \omega_{AB}\cZ^B{}_a,& \cZ^A{}_a &= \omega^{AB}\cZ_B{}_a,& \omega^{AC}\omega_{CB}&=\delta^A{}_B,
\end{align}
although $\om$ is not invariant under $SU(4)$.

When we go from the $\cN=5$ theory to the $\cN=6$ theory we will see 
that all occurrences of $\omega$ either disappear due to the antisymmetry condition on the triple product,
or combine into the $SU(4)$ invariant tensor $\epsilon$, according to the conventions
\begin{align}
\varepsilon^{ABCD}=-\om^{AB}\om^{CD}+\om^{AC}\om^{BD}-\om^{AD}\om^{BC},\nn\\
\label{epsid2}
\varepsilon_{ABCD}=-\om_{AB}\om_{CD}+\om_{AC}\om_{BD}-\om_{AD}\om_{BC}.
\end{align}
Replacing $A,B,C$ in the latter equation with $E,F,G$, and contracting the two equations we get the identity 
\cite{Hosomichi:2008jb}
\begin{align} \label{epsid3}
6\de^A{}_{[E} \de^B{}_F \de^C{}_{G]}=
-3\de^A{}_{[E} \om^{BC}\om_{FG]}-3\de^B{}_{[E} \om^{CA}\om_{FG]}-3\de^C{}_{[E} \om^{AB}\om_{FG]},
\end{align}
which we will use later.

We consider the matter fields $\cZ_A{}^a$ and $\Psi^{Aa}$ in the $\cN=5$ theory as components of 
elements $\cZ_A=\cZ_A{}^a T_a$ and $\Psi^A=\Psi^{Aa} T_a$ in an anti-Lie triple system, which is the odd subspace $\cG_{(1)}$ of a basic Lie superalgebra. Then $\cZ_A$ and $\Psi^{A}$ give rise to elements $Z_A$ and $\psi^A$ in an $\cN=5$ three-algebra, which is the subspace $\cG_{-1}$. The relation between the elements in the anti-Lie triple system and the $\cN=5$ three-algebra
is given by
\begin{align} \label{realitetsvillkor}
\cZ^A{}_\al &= Z^A{}_\al,     & \Psi_A{}_\al &= \psi_A{}_\al,\nn\\
\cZ_A{}^\al &= \bar Z_A{}^\al, & \Psi^A{}^\al &= -\psi^A{}^\al.
\end{align}
In the $\cN=6$ theory $\bar Z_A{}^\al$ is the complex conjugate of $Z^A{}_a$, and $\psi_A{}_\al$ is the complex conjugate of
$\psi^A{}^\al$,
\begin{align}
\bar Z_A{}^\al&=(Z^A{}_a)^\ast, & (\psi_A{}_\al)&=(\psi^A{}^\al)^\ast.
\end{align}
Therefore the relations (\ref{realitetsvillkor}) imply the reality conditions
\begin{align}
(\cZ^A{}_a)^\ast &= \cZ_A{}^a, &  \Psi_A{}_a = - (\Psi^A{}^a)^\ast.
\end{align}
The minus sign here 
does not agree with the constraints in 
\cite{Chen:2009cwa}, but
its importance was stressed in \cite{Bagger:2010zq}, so we believe that it is correct.

Beside the matter fields, the $\cN=5$ theory also contains a Chern-Simons gauge field $\cA_\mu{}^{ab}$ which is symmetric in the upper indices. We consider it as the components of an element $\cA_\mu$ of $\cG_{(0)}$, the even subalgebra of the Lie superalgebra $\cG$, which is generated 
by the basis elements $T_a$ of the odd subspace,
\begin{align} \label{chernsimonsdef}
\cA_\mu =\cA_{\mu}{}^{ab}\{T_a,T_b\}.
\end{align}
In other words, $\cG_{(0)}$ (which is an ordinary Lie algebra) is the gauge algebra of the theory. 
If $\cG$ is 5-graded with $\cG_{\pm2} \neq 0$, then $\cG_{(0)}$ is 3-graded, as can be seen in 
(\ref{evenoddparts}).
When we split the latin indices in (\ref{chernsimonsdef}) into greek ones, we get one term for each of the 
three subspaces (but one of them comes with a factor of two, due to the symmetry),
\begin{align}
\cA_\mu &=\cA_{\mu}{}^{\al \be}\{T_\al,T_\be\}+
2\cA_{\mu}{}^{\al \dot\be}\{T_\al,T_{\dot\be}\}
+\cA_{\mu}{}^{\dot\al \dot\be}\{T_{\dot\al},T_{\dot\be}\}\nn\\
&=\cA_{\mu}{}^{\al \be}\{T_\al,T_\be\}-
2\cA_{\mu}{}^{\al}{}_\be\{T_\al,T^{\be}\}
+\cA_{\mu}{}_{\al\be}\{T^{\al},T^{\be}\}\nn\\
&= \tfrac12A_{\mu}{}^{\al \be}\{T_\al,T_\be\}-
A_{\mu}{}^{\al}{}_\be\{T_\al,T^{\be}\}
+\tfrac12A_{\mu}{}_{\al\be}\{T^{\al},T^{\be}\},
\end{align}
where we in the last step have set
\begin{align} \label{aa-identification}
A_{\mu}{}^{\al\be}&=2\cA_{\mu}{}^{\al\be},&A_{\mu}{}^{\al}{}_\be&=2\cA_{\mu}{}^{\al}{}_\be,&
A_{\mu}{}_{\al\be}&=2\cA_{\mu}{}_{\al\be},
\end{align}
which implies that also $A_{\mu}{}^{\al\be}$ and $A_{\mu}{}_{\al\be}$ are symmetric in the two indices.

In addition to the Chern-Simons term, the gauge field also appears in the covariant derivative,
\begin{align}
D_\mu \cZ^A{}_d &= \partial_\mu \cZ^A{}_d-\cA_\mu{}^{ab}f_{abd}{}^c \cZ^A{}_c,\nn\\
D_\mu \cZ_A{}^d &= \partial_\mu \cZ_A{}^d+\cA_\mu{}^{ab}f_{abc}{}^d \cZ_A{}^c.
\end{align}
These equations imply
\begin{align} 
D_\mu Z^A{}_\de &= \partial_\mu Z^A{}_\de + A_\mu{}^\al{}_\be f^\be{}_\al{}^\ga{}_\de Z^A{}_\ga
-\om^{AB}A_\mu{}_{\al\be}f^\al{}_\de{}^\be{}_\ga \bar Z_B{}^\ga,\nn\\
D_\mu \bar Z_A{}^\de &= \partial_\mu \bar Z_A{}^\de + A_\mu{}^\al{}_\be f^\be{}_\al{}^\de{}_\ga \bar Z_A{}^\ga
+\om_{AB}A_\mu{}^{\al\be}f^\ga{}_\al{}^\de{}_\be Z^B{}_\ga. \label{covder5}
\end{align}
Note that, unlike the $\cN=6$ theory, a gauge transformation can take $Z^A$ to $\bar Z_B$ and vice versa. The elements in the gauge algebra that are responsible for such transformations belong to $\cG_{\pm 2}$. When $\cG$ is 3-graded, 
(\ref{covder5}) reduces to the definition of the covariant derivative in the $\cN=6$ theory, 
\begin{align}
D_\mu Z^A{}_\de &= \partial_\mu Z^A{}_\de + A_\mu{}^\al{}_\be f^\be{}_\al{}^\ga{}_\de Z^A{}_\ga
\equiv \partial_\mu Z^A{}_\de - \tilde A_\mu{}^\ga{}_\de  Z^A{}_\ga,\nn\\
D_\mu \bar Z_A{}^\de &= \partial_\mu \bar Z_A{}^\de + A_\mu{}^\al{}_\be f^\be{}_\al{}^\de{}_\ga \bar Z_A{}^\ga
\equiv \partial_\mu \bar Z_A{}^\de - \tilde A_\mu{}^\de{}_\ga  \bar Z_A{}^\ga. \label{cover6}
\end{align}

We are now ready to write the two Lagrangians that we want to unify.
We write them in the notation that we have introduced above, with greek indices for three-algebras and latin indices for anti-Lie triple systems,
but otherwise as they were given in \cite{Chen:2009cwa}.

Each of the Lagrangians consists of kinetic terms for the matter fields, a Yukawa coupling, a Chern-Simons term and the potential. 
We write this as
\begin{align}
\cL{}&=\cL_{\text{kin}}+\cL_{\text{Y}}+\cL_{\text{CS}}+\cL_{\text{pot}}.
\end{align}
Then the $\cN=5$ Lagrangian is
\begin{align}
\cL_{\text{kin}}&=
-\tfrac12D_\mu \cZ^A{}_a D^\mu \cZ_A{}^a+\tfrac12i\bar\Psi^A{}_a D_\mu \gamma^\mu \Psi_A{}^a,\nn\\
\cL_{\text{Y}}&=-\tfrac12{i}\omega^{AB}\omega^{CD}\omega_{de}f_{abc}{}^e(\cZ_A{}^a \cZ_B{}^c \bar \Psi_C{}^b \Psi_D{}^d
-2\cZ_A{}^a \cZ_D{}^c \bar \Psi_C{}^b \Psi_B{}^d),\nn\\
\cL_{\text{CS}}&=\tfrac12\varepsilon^{\mu\nu\rho}(\omega_{de}f_{abc}{}^e \cA_{\mu}{}^{ab}\partial_\nu \cA_\rho{}^{cd}
+\tfrac23 \omega_{fh}f_{abc}{}^g f_{gde}{}^h \cA_{\mu}{}^{ab}\cA_\nu{}^{cd} \cA_\rho{}^{ef}),\nn\\
\cL_{\text{pot}}&=-\tfrac1{60}(2f_{abc}{}^g f_{gdf}{}^e-9f_{cda}{}^g f_{gfb}{}^e+2f_{abd}{}^g f_{gcf}{}^e)
\cZ_A{}^f \cZ^A{}^a \cZ_B{}^b \cZ^B{}^c \cZ_C{}^d \cZ^C{}_e, 
\label{n5chen}
\end{align}
where $f_{abc}{}^d$ are the structure constants of an anti-Lie triple system,
and the $\cN=6$ Lagrangian is
\begin{align}
\cL_{\text{kin}}&=-D_\mu \bar Z_A{}^\al D^\mu Z^A{}_\al-i \bar\psi^{A\al}\ga^\mu D_\mu \psi_{A\al},\nn\\
\cL_{\text{Y}}&=-if^\al{}_\ga{}^\be{}_\de \bar\psi^{A\de}\psi_{A\al}Z^B{}_\be \bar Z_B{}^\ga+2if^\al{}_\ga{}^\be{}_\de\bar\psi^{A\de}\psi_{B\al}Z^B{}_\be
\bar Z_A{}^\ga\nn\\
&\quad\,-\tfrac{i}2\varepsilon_{ABCD}f^\al{}_\ga{}^\be{}_\de \bar \psi^{A\ga}\psi^{B\de}Z^C{}_\al Z^D{}_\be
-\tfrac{i}2\varepsilon^{ABCD}f^\ga{}_\al{}^\de{}_\be \bar \psi_{A\ga}\psi_{B\de}\bar Z_C{}^\al\bar Z_D{}^\be,\nn\\
\cL_{\text{CS}}&=\tfrac{1}2\varepsilon^{\mu\nu\rho}(f^\al{}_\ga{}^\be{}_\de A_\mu{}^\ga{}_\be\partial_\nu A_\rho{}^\de{}_\al
+\tfrac23 f^\al{}_\de{}^\ga{}_\eta f^{\eta}{}_\varphi{}^\varepsilon{}_\be A_{\mu}{}^{\be}{}_{\al}A_\nu{}^\de{}_\ga 
A_\rho{}^{\varphi}{}_\varepsilon),\nn\\
\cL_{\text{pot}}&=-\tfrac23(f^\al{}_\ga{}^\be{}_\de f^\varepsilon{}_\varphi{}^\de{}_\eta-\tfrac12 f^\varepsilon{}_\ga{}^\be{}_\de f^\al{}_\varphi{}^\de{}_\eta)
\bar Z_A{}^\ga Z^A{}_\varepsilon\bar Z_B{}^\varphi Z^B{}_\al\bar Z_C{}^\eta Z^C{}_\be, \label{n6chen}
\end{align}
where $f^\alpha{}_\beta{}^\gamma{}_\delta$ are the structure constants of an $\cN=6$ three-algebra.
Our claim is now that (\ref{n5chen}) and (\ref{n6chen}) can be unified into the single Lagrangian
\begin{align}
\cL_{\text{kin}}&=-D_\mu \bar Z_A{}^\al D^\mu Z^A{}_\al-i \bar\psi^{A\al}\ga^\mu D_\mu \psi_{A\al},\nn\\
\cL_{\text Y}&=
if^\al{}_\be{}^\ga{}_\de
(-Z^{A}{}_\al \bar Z_A{}^\be \bar \psi_B{}_\ga \psi^B{}^\de
-2 Z^A{}_\al \bar Z_{B}{}^\be \bar \psi_A{}_\ga \psi^B{}^\de)\nn\\
&\qquad\qquad+\tfrac{i}2f^\al{}_\be{}^\ga{}_\de Z^A{}_\al\bar\psi^C{}^\be Z^B{}_\ga \psi^{D}{}^\de (\om_{AB}\om_{CD}-2\om_{AD}\om_{CB})\nn\\
&\qquad\qquad+\tfrac{i}2f^\al{}_\be{}^\ga{}_\de \bar Z_B{}^\de\bar\psi_C{}_\al \bar Z_A{}^\be \psi_{D}{}_\ga (\om^{AB}\om^{CD}-2\om^{AD}\om^{CB})\nn\\
&\qquad\qquad+2if^\al{}_\be{}^\ga{}_\de
\bar\psi_C{}_{(\al}\bar Z_B{}^\be Z^A{}_{\ga)} \psi^D{}^\de(\de^B{}_A\de^C{}_D-2\om_{AD}\om^{CB}),\nn\\
\cL_{\text{CS}}&=\tfrac{1}2\varepsilon^{\mu\nu\rho}
(f^{\al}{}_\be{}^\ga{}_{\de}A_{\mu}{}_{\al}{}^{\be}\partial_{\nu}A_{\rho}{}_{\ga}{}^{\de}
+f^{\al}{}_{\be}{}^{\ga}{}_{\eta} f^{\eta}{}_{\de}{}^\varepsilon{}_\varphi 
A_{\mu}{}_{\al}{}^{\be}A_{\nu}{}_{\ga}{}^{\de} A_{\rho}{}_{\varepsilon}{}^{\varphi}
\nn\\
&\qquad\qquad\,-f^{\al}{}_\ga{}^\be{}_{\de}A_{\mu}{}_{\al\be}\partial_{\nu}A_{\rho}{}^{\ga\de}\nn\\
&\qquad\qquad\,- (f^\al{}_\ga{}^\be{}_\eta f^\eta{}_\varepsilon{}^\de{}_\varphi 
+2 f^\de{}_\ga{}^\be{}_\eta f^\eta{}_\varepsilon{}^\al{}_\varphi)
A_{\mu}{}_{\al\be}A_{\nu}{}^\ga{}_\de 
A_{\rho}{}^{\varepsilon\varphi})\nn\\
\cL_{\text{pot}}&=f^\al{}_\be{}^\ga{}_\eta f^\de{}_\varepsilon{}^\eta{}_\varphi
(-\om_{AB}\om^{DE}Z^A{}_\al \bar Z_D{}^\be Z^C{}_\ga \bar Z_E{}^\varepsilon Z^B{}_\de \bar Z_C{}^\varphi\nn\\
&\quad\,\,\,
-\om_{AB}\om^{DE}
Z^A{}_\al \bar Z_D{}^\be Z^B{}_\ga \bar Z_C{}^\varepsilon Z^C{}_\de \bar Z_E{}^\varphi
-Z^A{}_\al \bar Z_A{}^\be Z^C{}_\ga \bar Z_B{}^\varepsilon Z^B{}_\de \bar Z_C{}^\varphi),
\label{n5jp}
\end{align}
where $f^\alpha{}_\beta{}^\gamma{}_\delta$ are the structure constants of an $\cN=5$ three-algebra.
In the following two sections we will prove this claim.

\section{Derivation from anti-Lie triple systems}

In this section we show that (\ref{n5jp}) can be derived from (\ref{n5chen}). 
The general strategy is to split the latin indices into dotted and undotted greek ones, and then remove the 
dots with $\Om$, as explained in section \ref{Connection to 5-graded Lie superalgebras}. Alternatively, we can go from the anti-Lie triple system to the $\cN=5$ three-algebra, via the associated Lie superalgebra, by writing out the triple products and inner products, without using indices. As we will see in section \ref{potsec}, this is particularly convenient for the potential, where there are many indices to keep track of. But we start with the much simpler kinetic terms. 

\subsection{The kinetic terms}

When we replace the latin indices with greek ones in the kinetic terms, we pick up a factor of 2 because of the antisymmetry of $\Om$. For the fermions we furthermore pick up a minus sign since we have to raise and lower the contracted greek index before inserting 
(\ref{realitetsvillkor}). Thus when we take the kinetic terms $\cL_{\text{kin}}$ in (\ref{n5chen}) we have
\begin{align}
-\tfrac12 D_\mu \cZ^A{}_a D^\mu \cZ_A{}^a &= -\tfrac12\Om^{ab} D_\mu \cZ^A{}_a D^\mu\cZ_A{}_b
= -\Om^{\al\dot\be} D_\mu \cZ^A{}_\al D^\mu\cZ_A{}_{\dot\be}\nn\\
&=- D_\mu \cZ^A{}_\al D^\mu \cZ_A{}^\al= - D_\mu Z^A{}_\al D^\mu \bar Z_A{}^\al,
\end{align}
\begin{align}
\tfrac{i}2 \bar\Psi^A{}_a D_\mu \ga^\mu \Psi_A{}^a = i \bar\Psi^A{}_\al D_\mu \ga^\mu \Psi_A{}^\al
=-i \bar\Psi^A{}^\al D_\mu \ga^\mu \Psi_A{}_\al
=-i \bar\psi^A{}^\al D_\mu \ga^\mu \psi_A{}_\al,
\end{align}
and we end up with the kinetic terms $\cL_{\text{kin}}$ in (\ref{n5jp}).

\subsection{The Yukawa coupling}

We decompose the Yukawa coupling in (\ref{n5chen}) into two terms, $\cL_{\text{Y}}=\cL_{\text{Y}1}+\cL_{\text{Y}2}$, where
\begin{align}
\cL_{\text{Y}1}&=-\tfrac{i}2\om^{AB}\om^{CD}f_{abcd}\cZ_A{}^a\cZ_B{}^c\bar\Psi_C{}^b\Psi_D{}^d,\nn\\
\cL_{\text{Y}2}&=i\om^{AB}\om^{CD}f_{abcd}\cZ_A{}^a\cZ_D{}^c\bar\Psi_C{}^b\Psi_B{}^d.
\end{align}
The first term can be written
\begin{align}
\cL_{\text{Y}1}=-\tfrac{i}2\om^{AB}\om^{CD}
(f_{\dot\al\be\dot\ga\de}\cZ_A{}^{\dot\al}\cZ_B{}^{\dot\ga}\bar\Psi_C{}^{\be}\Psi_D{}^{\de}
+&f_{\al\dot\be\ga\dot\de}\cZ_A{}^{\al}\cZ_B{}^{\ga}\bar\Psi_C{}^{\dot\be}\Psi_D{}^{\dot\de}\nn\\
+2f_{\dot\al\be\ga\dot\de}\cZ_A{}^{\dot\al}\cZ_B{}^\ga\bar\Psi_C{}^{\be}\Psi_D{}^{\dot\de}
+&2f_{\dot\al\dot\be\ga\de}\cZ_A{}^{\dot\al}\cZ_B{}^\ga\bar\Psi_C{}^{\dot\be}\Psi_D{}^\de)\nn\\
=-\tfrac{i}2\om^{AB}\om^{CD}f^\al{}_\be{}^\ga{}_\de
(\cZ_{A\al}\cZ_{B\ga}\bar\Psi_C{}^{\be}\Psi_D{}^{\de}
+&\cZ_A{}^{\be}\cZ_B{}^{\de}\bar\Psi_C{}_{\al}\Psi_D{}_{\ga}\nn\\
-2\cZ_{A\al}\cZ_B{}^\be\bar\Psi_{C\ga}\Psi_D{}^\de-&2\cZ_{A\ga}\cZ_B{}^\be\bar\Psi_{C\al}\Psi_D{}^\de\nn\\
+&2\cZ_{A\al}\cZ_B{}^{\de}\bar\Psi_C{}^{\be}\Psi_{D\ga})\nn\\
=-\tfrac{i}2\om^{AB}\om^{CD}f^\al{}_\be{}^\ga{}_\de
(\cZ_{A\al}\cZ_{B\ga}\bar\Psi_C{}^{\be}\Psi_D{}^{\de}
+&\cZ_A{}^{\be}\cZ_B{}^{\de}\bar\Psi_C{}_{\al}\Psi_D{}_{\ga}\nn\\
-2\cZ_{A\al}\cZ_B{}^\be\bar\Psi_{C\ga}\Psi_D{}^\de-&4\cZ_{A\ga}\cZ_B{}^\be\bar\Psi_{C\al}\Psi_D{}^\de),
\end{align}
where we in the second step we have replaced dotted indices downstairs with undotted upstairs (and vice versa), and used 
(\ref{altscomp2}).
In the last step we have used (\ref{altscomp3}),
and the antisymmetry of $\om^{CD}$. For the second term we get 
likewise
\begin{align}
\cL_{\text{Y}2}=i\om^{AB}\om^{CD}f^\al{}_\be{}^\ga{}_\de(&-\cZ_{A\al}\cZ_D{}^\be\bar\Psi_{C\ga}\Psi_B{}^\de
-\cZ_{A}{}^\be\cZ_D{}_\ga\bar\Psi_{C}{}^{\de}\Psi_{B\al}\nn\\
&-\cZ_{A\ga}\cZ_D{}^\be\bar\Psi_{C\al}\Psi_B{}^\de
-\cZ_{A}{}^\be\cZ_D{}_\al\bar\Psi_{C}{}^{\de}\Psi_{B\ga}\nn\\
&+\cZ_{A\al}\cZ_D{}^\de\bar\Psi_{C}{}^\be\Psi_{B\ga}
+\cZ_{A}{}^\be\cZ_D{}_\ga\bar\Psi_{C}{}_\al\Psi_{B}{}^{\de}\nn\\
&+\cZ_{A\al}\cZ_D{}_\ga\bar\Psi_{C}{}^\be\Psi_{B}{}^{\de}
+\cZ_{A}{}^{\be}\cZ_D{}^\de\bar\Psi_{C}{}_\al\Psi_{B}{}_{\ga})\nn\\
=i\om^{AB}\om^{CD}f^\al{}_\be{}^\ga{}_\de(&-2\cZ_{A\al}\cZ_D{}^\be\bar\Psi_{C\ga}\Psi_B{}^\de
\nn\\
&-2\cZ_{A\ga}\cZ_D{}^\be\bar\Psi_{C\al}\Psi_B{}^\de
\nn\\
&+2\cZ_{A\al}\cZ_D{}^\de\bar\Psi_{C}{}^\be\Psi_{B\ga}
\nn\\
&+\cZ_{A\al}\cZ_D{}_\ga\bar\Psi_{C}{}^\be\Psi_{B}{}^{\de}
+\cZ_{A}{}^{\be}\cZ_D{}^\de\bar\Psi_{C}{}_\al\Psi_{B}{}_{\ga}),
\end{align}
and the two terms add up to
\begin{align}
\cL_{\text Y}&=
if^\al{}_\be{}^\ga{}_\de(\cZ_{A}{}_\al \cZ_B{}^\be \bar \Psi_C{}_\ga \Psi_D{}^\de
+\bar\Psi_C{}_\al \cZ_B{}^\be \cZ_{A}{}_\ga \Psi_D{}^\de\nn\\
&\qquad\qquad-\tfrac12\cZ_{A}{}_\al \bar \Psi_C{}^\be \cZ_B{}_\ga  \Psi_D{}^\de
-\tfrac12\bar \Psi_C{}_\al \cZ_A{}^\be \Psi_{D}{}_\ga \cZ_B{}^\de\nn\\
&\qquad\qquad-\bar\Psi_{C}{}_\al \cZ_A{}^\be \cZ_B{}_\ga \Psi_D{}^\de)(\om^{AB}\om^{CD}-2\om^{AD}\om^{CB}).
\end{align}
Inserting (\ref{realitetsvillkor}) this becomes
\begin{align} \label{three-alg-yukawa}
\cL_{\text Y}&=
if^\al{}_\be{}^\ga{}_\de
(-Z^{A}{}_\al \bar Z_A{}^\be \bar \psi_B{}_\ga \psi^B{}^\de
-2 Z^A{}_\al \bar Z_{B}{}^\be \bar \psi_A{}_\ga \psi^B{}^\de)\nn\\
&\qquad\qquad+\tfrac{i}2f^\al{}_\be{}^\ga{}_\de Z^A{}_\al\bar\psi^C{}^\be Z^B{}_\ga \psi^{D}{}^\de (\om_{AB}\om_{CD}-2\om_{AD}\om_{CB})\nn\\
&\qquad\qquad+\tfrac{i}2f^\al{}_\be{}^\ga{}_\de \bar Z_B{}^\de\bar\psi_C{}_\al \bar Z_A{}^\be \psi_{D}{}_\ga (\om^{AB}\om^{CD}-2\om^{AD}\om^{CB})\nn\\
&\qquad\qquad+2if^\al{}_\be{}^\ga{}_\de
\bar\psi_C{}_{(\al}\bar Z_B{}^\be Z^A{}_{\ga)} \psi^D{}^\de(\de^B{}_A\de^C{}_D-2\om_{AD}\om^{CB}),
\end{align}
which is again what we have in (\ref{n5jp}).

\subsection{The Chern-Simons term}
We proceed to the Chern-Simons term,
\begin{align}
\cL_{\text{CS}}&=\tfrac12\varepsilon^{\mu\nu\rho}(\omega_{de}f_{abc}{}^e \cA_{\mu}{}^{ab}\partial_\nu \cA_\rho{}^{cd}
+\tfrac23 \omega_{fh}f_{abc}{}^g f_{gde}{}^h \cA_{\mu}{}^{ab}\cA_\nu{}^{cd} \cA_\rho{}^{ef}).
\end{align}
Up to a total derivative we have
\begin{align}
f_{abcd}\cA_{[\mu}{}^{ab}\partial_{\nu}\cA_{\rho]}{}^{cd}
&=2f_{\dot\al\dot\be\ga\de}\cA_{[\mu}{}^{\dot\al\dot\be}\partial_{\nu}\cA_{\rho]}{}^{\ga\de}
+4f_{\dot\al\be\dot\ga\de}\cA_{[\mu}{}^{\dot\al\be}\partial_{\nu}\cA_{\rho]}{}^{\dot\ga\de}\nn\\
&=-4f_{\dot\al\ga\dot\be\de}\cA_{[\mu}{}^{\dot\al\dot\be}\partial_{\nu}\cA_{\rho]}{}^{\ga\de}
+4f_{\dot\al\be\dot\ga\de}\cA_{[\mu}{}^{\dot\al\be}\partial_{\nu}\cA_{\rho]}{}^{\dot\ga\de}\nn\\
&=-4f^{\al}{}_{\ga}{}^{\be}{}_{\de}\cA_{[\mu|}{}_{\al\be}\partial_{|\nu}\cA_{\rho]}{}^{\ga\de}
+4f^{\al}{}_{\be}{}^{\ga}{}_{\de}\cA_{[\mu|}{}_{\al}{}^{\be}\partial_{\nu}\cA_{\rho]}{}_{\ga}{}^{\de}
\end{align}
for the first half of the Chern-Simons term, and
\begin{align}
f_{abc}{}^h f_{hdef} \cA_{[\mu}{}^{ab}\cA_\nu{}^{cd} \cA_{\rho]}{}^{ef}&=
2f_{\dot\al\dot\be\ga}{}^{\dot\eta} f_{\dot\eta\dot\de\varepsilon\varphi} 
\cA_{[\mu}{}^{\dot\al\dot\be}\cA_\nu{}^{\ga\dot\de} \cA_{\rho]}{}^{\varepsilon\varphi}\nn\\
&\quad\,+8f_{\al\be\dot\ga}{}^{\eta} f_{\eta\dot\de\dot\varepsilon\varphi} 
\cA_{[\mu}{}^{\al\be}\cA_\nu{}^{\dot\ga\dot\de} \cA_{\rho]}{}^{\dot\varepsilon\varphi}\nn\\
&\quad\,+8f_{\dot\al\be\dot\ga}{}^{\dot\eta} f_{\dot\eta\de\dot\varepsilon\varphi} 
\cA_{[\mu}{}^{\dot\al\be}\cA_\nu{}^{\dot\ga\de} \cA_{\rho]}{}^{\dot\varepsilon\varphi}\nn\\
&=-8 f^\al{}_\ga{}^\be{}_\eta f^\eta{}_\varepsilon{}^\de{}_\varphi \cA_{[\mu|}{}_{\al\be}\cA_{|\nu|}{}^\ga{}_\de 
\cA_{|\rho]}{}^{\varepsilon\varphi}\nn\\
&\quad\,-16 f^\eta{}_\al{}^\ga{}_\be f^\de{}_\eta{}^\varepsilon{}_\varphi \cA_{[\mu}{}^{\al\be}\cA_{\nu|}{}_{\ga\de}
\cA_{|\rho]}{}^\varphi{}_\varepsilon\nn\\
&\quad\,+8f^{\al}{}_{\be}{}^{\ga}{}_{\eta} f^{\eta}{}_{\de}{}^\varepsilon{}_\varphi 
\cA_{[\mu|}{}_{\al}{}^{\be}\cA_{|\nu|}{}_{\ga}{}^{\de} \cA_{|\rho]}{}_{\varepsilon}{}^{\varphi}
\end{align}
for the second. Inserting (\ref{aa-identification}) this gives
\begin{align}
\cL_{\text{CS}}&=\tfrac12\varepsilon^{\mu\nu\rho}(\omega_{de}f_{abc}{}^e \cA_{\mu}{}^{ab}\partial_\nu \cA_\rho{}^{cd}
+\tfrac23 \omega_{fh}f_{abc}{}^g f_{gde}{}^h \cA_{\mu}{}^{ab}\cA_\nu{}^{cd} \cA_\rho{}^{ef})\nn\\
&=\tfrac{1}2\varepsilon^{\mu\nu\rho}
(-f^{\al}{}_\ga{}^\be{}_{\de}A_{\mu}{}_{\al\be}\partial_{\nu}A_{\rho}{}^{\ga\de}
+f^{\al}{}_\be{}^\ga{}_{\de}A_{\mu}{}_{\al}{}^{\be}\partial_{\nu}A_{\rho}{}_{\ga}{}^{\de}\nn\\
&\qquad\qquad\,- (f^\al{}_\ga{}^\be{}_\eta f^\eta{}_\varepsilon{}^\de{}_\varphi 
+2 f^\de{}_\ga{}^\be{}_\eta f^\eta{}_\varepsilon{}^\al{}_\varphi)
A_{\mu}{}_{\al\be}A_{\nu}{}^\ga{}_\de 
A_{\rho}{}^{\varepsilon\varphi}\nn\\
&\qquad\qquad\,+f^{\al}{}_{\be}{}^{\ga}{}_{\eta} f^{\eta}{}_{\de}{}^\varepsilon{}_\varphi 
A_{\mu}{}_{\al}{}^{\be}A_{\nu}{}_{\ga}{}^{\de} A_{\rho}{}_{\varepsilon}{}^{\varphi}),
\end{align}
so also this part of the Lagrangian (\ref{n5chen}) agrees with (\ref{n5jp}). The only remaining part is now the potential.

\subsection{The potential} \label{potsec}

In (\ref{n5chen}) the $\cN=5$ potential is (following \cite{Chen:2009cwa}) written as a sum of three terms,
$\cL_{\text{pot}}=V_1+V_2+V_3$, where
\begin{align}
V_1&=-\tfrac1{30}f_{abc}{}^g f_{gdf}{}^e\cZ_A{}^f \cZ^A{}^a \cZ_B{}^b \cZ^B{}^c \cZ_C{}^d \cZ^C{}_e,\nn\\
V_2&=\tfrac3{20}f_{cda}{}^g f_{gfb}{}^e\cZ_A{}^f \cZ^A{}^a \cZ_B{}^b \cZ^B{}^c \cZ_C{}^d \cZ^C{}_e,\nn\\
V_3&=-\tfrac1{30}f_{abd}{}^g f_{gcf}{}^e \cZ_A{}^f \cZ^A{}^a \cZ_B{}^b \cZ^B{}^c \cZ_C{}^d \cZ^C{}_e.
\label{potentialterms}
\end{align}
Instead of splitting the latin indices into dotted and undotted greek ones, it can be instructive to write each term in the Lie superalgebra language.
Using the invariance the first term can be written
\begin{align}
V_1&=-\tfrac1{30}f_{abc}{}^g f_{gdf}{}^e \cZ_A{}^f \cZ^A{}^a \cZ_B{}^b \cZ^B{}^c \cZ_C{}^d \cZ^C{}_e\nn\\
&=-\tfrac1{30}\big( \big[\big\{[\{\cZ^A,\cZ_B\},\cZ^B],\cZ_C\big\},\cZ_A\big] \big| \cZ^C\big) \nn\\
&=-\tfrac1{30}\big( [\{\cZ^A,\cZ_B\},\cZ^B] \big| [\cZ_C,\{\cZ_A,\cZ^C\}]\big),
\end{align}
and in the same way
\begin{align}
V_2&=\tfrac3{20}\big( [\{\cZ^B,\cZ_C\},\cZ^A] \big| [\cZ_A,\{\cZ_B,\cZ^C\}]\big),\nn\\
V_3&=-\tfrac1{30} \big( [\{\cZ^A,\cZ_B\},\cZ_C] \big| [\cZ^B,\{\cZ_A,\cZ^C\}]\big).
\end{align}
We can then go from the Lie superalgebra language to the $\cN=5$ three-algebra language
by writing
\begin{align}
\cZ^A = \cZ^A{}^a T_a &= \cZ^A{}^\al T_\al + \cZ^A{}^{\dot\al} T_{\dot\al}= \cZ^A{}^\al T_\al - \cZ^A{}_{\al} T^{\al}
=\tau(\bar Z^A)-Z^A,
\end{align}
and then using the definition (\ref{fromlsaton5}). To get simpler expressions,
we introduce the following shorthand notation,
\begin{align}
I_{ABC}&=\la (Z^A \bar Z^B Z^C)|(\bar Z_A Z_B \bar Z_C)\ra, &
I_{CBA}&=\la (Z^A \bar Z^B Z^C)|(\bar Z_C Z_B \bar Z_A)\ra,\nn\\
I_{BCA}&=\la (Z^A \bar Z^B Z^C)|(\bar Z_B Z_C \bar Z_A)\ra, &
I_{ACB}&=\la (Z^A \bar Z^B Z^C)|(\bar Z_A Z_C \bar Z_B)\ra,\nn\\
I_{CAB}&=\la (Z^A \bar Z^B Z^C)|(\bar Z_C Z_A \bar Z_B)\ra, &
I_{BAC}&=\la (Z^A \bar Z^B Z^C)|(\bar Z_B Z_A \bar Z_C)\ra,\nn
\end{align}
\begin{align}
I_{LL}&=\la (Z^C \bar Z^A Z_A)|(\bar Z_C Z^B \bar Z_B)\ra, &
I_{RL}&=\la (Z^A \bar Z_A Z^C)|(\bar Z_C Z^B \bar Z_B)\ra,\nn\\
I_{LC}&=\la (Z^C \bar Z^A Z_A)|(\bar Z^B Z_C \bar Z_B)\ra, &
I_{RC}&=\la (Z^A \bar Z_A Z^C)|(\bar Z^B Z_C \bar Z_B)\ra,\nn\\
I_{LR}&=\la (Z^C \bar Z^A Z_A)|(\bar Z^B Z_B \bar Z_C)\ra, &
I_{RR}&=\la (Z^A \bar Z_A Z^C)|(\bar Z^B Z_B \bar Z_C)\ra,\nn
\end{align}
\begin{align}
I_{CL}&=\la (Z^A \bar Z^C Z_A)|(\bar Z_C Z^B \bar Z_B)\ra, \nn\\
I_{CC}&=\la (Z^A \bar Z^C Z_A)|(\bar Z^B Z_C \bar Z_B)\ra, \nn\\
I_{CR}&=\la (Z^A \bar Z^C Z_A)|(\bar Z^B Z_B \bar Z_C)\ra. \label{i-termer}
\end{align}
In component form we have for example 
\begin{align}
I_{RR}=
-f^\al{}_\be{}^\ga{}_\eta f^\de{}_\varepsilon{}^\eta{}_\varphi
Z^A{}_\al \bar Z_A{}^\be Z^C{}_\ga \bar Z_B{}^\varepsilon Z^B{}_\de \bar Z_C{}^\varphi.
\end{align}
Taking the complex conjugate we get
\begin{align} \label{ccbca}
I_{BCA}{}^\ast &= \la (Z^A \bar Z^B Z^C)|(\bar Z_B Z_C \bar Z_A)\ra^\ast
= \la (\bar Z_A Z_B \bar Z_C)|(Z^B \bar Z^C Z^A)\ra\nn\\
&=\la(Z^B \bar Z^C Z^A)|(\bar Z_A Z_B \bar Z_C)\ra
= \la(Z^A \bar Z^B Z^C)|(\bar Z_C Z_A \bar Z_B)\ra=I_{CAB},
\end{align}
and in the same way 
$\xeight{}^\ast=\xten$, whereas each of the other expressions in (\ref{i-termer}) is its own conjugate.

By contracting the identities defining an $\cN=5$ three-algebra with six scalars, where the 
three-algebra indices and R-symmetries are combined in differents ways, we get a number of
non-trivial identities relating the expressions (\ref{i-termer}) to each other.
From (\ref{N=5ta1}) we get
\begin{align}
I_{CC}-I_{ACB}-I_{LC}+I_{CAB}&=0,&\xtwelvehalf&=\xtwelve,\nn\\
I_{CC}-I_{ACB}-I_{CL}+I_{BCA}&=0,&\xninehalf&=\xnine. \label{deforsta}
\end{align}
and from (\ref{N=5ta2}) follow another two identities,
\begin{align} 
\xtwelvehalf&=
\xeight+2\xfour+\xsix+2\xtwo+\xthree,\nn\\
\label{ytterligare2}
\xthirteen&=\xnine+\xone+\xfive.
\end{align}
Finally we also get an identity from (\ref{epsid3}) contracted with, for example, 
\begin{align}
\la(Z^E \bar Z^F Z^G)|(\bar Z_A Z_B \bar Z_C)\ra.
\end{align}
This identity involves all the 15 expressions and reads
\begin{align}
6I_{[ABC]}&= \label{15terms}
I_{LL}-I_{CL}+I_{RL}
-I_{LC}+I_{CC}-I_{RC} 
+I_{LR}-I_{CR}+I_{RR}.
\end{align}
The potential should be real, so we are interested in linear combinations that are real.
Therefore we write $J$ for the real part of $I$, for example
\begin{align}
J_{ABC}&=I_{ABC}, & J_{BCA}&=\frac12(I_{BCA}+I_{BCA}{}^\ast)=\frac12(I_{BCA}+I_{CAB}).
\end{align}
Since $I_{BCA}{}^\ast=I_{CAB}$ and $\xeight{}^\ast=\xten$, we have
\begin{align} \label{bcacablccl}
J_{BCA}&=J_{CAB},&J_{LC}&=J_{CL}.
\end{align}
The identities (\ref{deforsta}), (\ref{ytterligare2}) and (\ref{15terms}) of course hold also with $I$ replaced by $J$ everywhere. Together with 
(\ref{bcacablccl}) they give
\begin{align} 
J_{CC}&=-2J_{BAC}-4J_{CAB}+J_{CR},\nn\\
J_{LR}&=J_{RL}=J_{ABC}-J_{CBA}+J_{RR},\nn\\
J_{LC}&=J_{CL}=-J_{ACB}-2J_{BAC}-3J_{CAB}+J_{CR},\nn\\
J_{LL}&=3J_{ABC}-3J_{ACB}-3J_{BAC}+J_{CBA}+3J_{CR}-3J_{RR}.  \label{n=5relationer}
\end{align}
We can now write the potential as a linear combination of the terms on the right hand side of 
(\ref{n=5relationer}). Since this calculation is quite tedious, and since we have already demonstrated the general method for the other parts of the Lagrangian, we just present the result for the three terms (\ref{potentialterms}) of the potential,
\begin{align}
V_1&=-\tfrac2{30}(-22J_{BAC}-16J_{CAB}-16J_{ACB}+8J_{CBA}+16J_{ABC}+15J_{CR}-15J_{RR}),\nn\\
V_2&=\tfrac3{10}(2J_{BAC}-4J_{CAB}-4J_{ACB}+2J_{CBA}+4J_{ABC}),\nn\\
V_3&=-\tfrac2{30}(J_{BAC}-2J_{CAB}-2J_{ACB}+J_{CBA}+2J_{ABC}),
\end{align}
and we see that two of them are in fact linearly dependent, $V_2=-9V_3$.
Adding up the three terms we get the potential 
\begin{align}
\cL_{\text{pot}}&=V_1+V_2+V_3
=2J_{BAC}-J_{CR}+J_{RR}=2I_{BAC}-I_{CR}+I_{RR},
\end{align}
where we have used that $I_{BAC}$, $I_{CR}$ and $I_{RR}$ are real, so that they coincide with 
$J_{BAC}$, $J_{CR}$ and $J_{RR}$. Reinserting (\ref{i-termer}) we finally obtain
\begin{align}
\cL_{\text{pot}}
&=2I_{BAC}-I_{CR}+I_{RR}\nn\\
&=2 \la (Z^A \bar Z^B Z^C)|(\bar Z_B Z_A \bar Z_C)\ra\nn\\
&\quad\,-  \la (Z^A \bar Z^C Z_A)|(\bar Z^B Z_B \bar Z_C)\ra
+ \la (Z^A \bar Z_A Z^C)|(\bar Z^B Z_B \bar Z_C)\ra\nn\\
&= f^\al{}_\be{}^\ga{}_\eta f^\de{}_\varepsilon{}^\eta{}_\varphi
(-\om_{AB}\om^{DE}Z^A{}_\al \bar Z_D{}^\be Z^C{}_\ga \bar Z_E{}^\varepsilon Z^B{}_\de \bar Z_C{}^\varphi\nn\\
&\quad\,\,\,
-\om_{AB}\om^{DE}
Z^A{}_\al \bar Z_D{}^\be Z^B{}_\ga \bar Z_C{}^\varepsilon Z^C{}_\de \bar Z_E{}^\varphi
-Z^A{}_\al \bar Z_A{}^\be Z^C{}_\ga \bar Z_B{}^\varepsilon Z^B{}_\de \bar Z_C{}^\varphi),
\end{align}
and we have thus arrived at the expression for the potential in 
(\ref{n5jp}).

\section{From $\cN=5$ to $\cN=6$}

In the preceding section we started with the $\cN=5$ Lagrangian (\ref{n5chen}) and showed that it leads to our new Lagrangian (\ref{n5jp}) when we go from anti-Lie triple systems to $\cN=5$ three-algebras.
In this section we will instead start with (\ref{n5jp}) and show that it 
reduces to the $\cN=6$ Lagrangian (\ref{n6chen}) when we impose the antisymmetry condition on the triple product,
so that the $\cN=5$ three-algebra reduces to an $\cN=6$ three-algebra.

Again we start with the kinetic terms $\cL_{\text{kin}}$. They have already the same form in (\ref{n5jp}) as in (\ref{n6chen}),
so we only have to show that the covariant derivative in (\ref{n5jp}) reduces to the one in (\ref{n6chen}) when we impose the antisymmetry condition.
But this we already did when we defined the covariant derivatives in 
(\ref{covder5}) and (\ref{cover6}), so we can conclude that the kinetic terms agree.

We proceed to the Yukawa coupling, which in (\ref{n5jp}) has the form
\begin{align} \label{three-alg-yukawa2}
\cL_{\text Y}&=
if^\al{}_\be{}^\ga{}_\de
(-Z^{A}{}_\al \bar Z_A{}^\be \bar \psi_B{}_\ga \psi^B{}^\de
-2 Z^A{}_\al \bar Z_{B}{}^\be \bar \psi_A{}_\ga \psi^B{}^\de)\nn\\
&\qquad\qquad+\tfrac{i}2f^\al{}_\be{}^\ga{}_\de Z^A{}_\al\bar\psi^C{}^\be Z^B{}_\ga \psi^{D}{}^\de (\om_{AB}\om_{CD}-2\om_{AD}\om_{CB})\nn\\
&\qquad\qquad+\tfrac{i}2f^\al{}_\be{}^\ga{}_\de \bar Z_B{}^\de\bar\psi_C{}_\al \bar Z_A{}^\be \psi_{D}{}_\ga (\om^{AB}\om^{CD}-2\om^{AD}\om^{CB})\nn\\
&\qquad\qquad+2if^\al{}_\be{}^\ga{}_\de
\bar\psi_C{}_{(\al}\bar Z_B{}^\be Z^A{}_{\ga)} \psi^D{}^\de(\de^B{}_A\de^C{}_D-2\om_{AD}\om^{CB}).
\end{align}
When we impose the antisymmetry condition on the triple product 
the last line vanishes, and we can rewrite the preceding two lines,
\begin{align} \label{three-alg-yukawa3}
\cL_{\text Y}&=if^\al{}_\be{}^\ga{}_\de
(-Z^{A}{}_\al \bar Z_A{}^\be \bar \psi_B{}_\ga \psi^B{}^\de
-2 Z^A{}_\al \bar Z_{B}{}^\be \bar \psi_A{}_\ga \psi^B{}^\de)\nn\\
&\quad\,+\tfrac{i}2 
f^\al{}_\ga{}^\be{}_\de \bar \psi^{A\ga}\psi^{B\de}Z^C{}_\al Z^D{}_\be
(\om_{AB}\om_{CD}-\om_{AD}\om_{CB}+\om_{BD}\om_{CA})\nn\\
&\quad\,+\tfrac{i}2  f^\ga{}_\al{}^\de{}_\be \bar \psi_{A\ga}\psi_{B\de}\bar Z_C{}^\al\bar Z_D{}^\be 
(\om^{AB}\om^{CD}-\om^{AD}\om^{CB}+\om^{BD}\om^{CA}).
\end{align}
Using the identities (\ref{epsid2})
we then get
\begin{align} \label{three-alg-yukawa3}
\cL_{\text{Y}}&=-if^\al{}_\ga{}^\be{}_\de \bar\psi^{A\de}\psi_{A\al}Z^B{}_\be \bar Z_B{}^\ga+2if^\al{}_\ga{}^\be{}_\de\bar\psi^{A\de}\psi_{B\al}Z^B{}_\be
\bar Z_A{}^\ga\nn\\
&\quad\,-\tfrac{i}2\varepsilon_{ABCD}f^\al{}_\ga{}^\be{}_\de \bar \psi^{A\ga}\psi^{B\de}Z^C{}_\al Z^D{}_\be
-\tfrac{i}2\varepsilon^{ABCD}f^\ga{}_\al{}^\de{}_\be \bar \psi_{A\ga}\psi_{B\de}\bar Z_C{}^\al\bar Z_D{}^\be,
\end{align}
which agrees with the $\cN=6$ Lagrangian (\ref{n6chen}).

The Chern-Simons term in (\ref{n5jp}) has the form
\begin{align} \label{csterm2}
\cL_{\text{CS}}&=\tfrac{1}2\varepsilon^{\mu\nu\rho}
(f^{\al}{}_\be{}^\ga{}_{\de}A_{\mu}{}_{\al}{}^{\be}\partial_{\nu}A_{\rho}{}_{\ga}{}^{\de}
+f^{\al}{}_{\be}{}^{\ga}{}_{\eta} f^{\eta}{}_{\de}{}^\varepsilon{}_\varphi 
A_{\mu}{}_{\al}{}^{\be}A_{\nu}{}_{\ga}{}^{\de} A_{\rho}{}_{\varepsilon}{}^{\varphi}
\nn\\
&\qquad\qquad\,-f^{\al}{}_\ga{}^\be{}_{\de}A_{\mu}{}_{\al\be}\partial_{\nu}A_{\rho}{}^{\ga\de}\nn\\
&\qquad\qquad\,- (f^\al{}_\ga{}^\be{}_\eta f^\eta{}_\varepsilon{}^\de{}_\varphi 
+2 f^\de{}_\ga{}^\be{}_\eta f^\eta{}_\varepsilon{}^\al{}_\varphi)
A_{\mu}{}_{\al\be}A_{\nu}{}^\ga{}_\de 
A_{\rho}{}^{\varepsilon\varphi}).
\end{align}
Except for the Chern-Simons term for the $\cN=6$ gauge field $A_\mu{}_\al{}^\be$, which we want to keep,
(\ref{csterm2}) also contains terms involving the additional gauge fields $A_\mu{}_{\al\be}$ and 
$A_\mu{}^{\al\be}$ corresponding to the $\cG_{\pm 2}$ subspaces in the Lie superalgebra. But as in the covariant derivatives,
these fields are always contracted with the structure constants $f^{\al}{}_{\be}{}^\ga{}_\de$ of the $\cN=5$ three-algebra.
When the triple product is antisymmetric in the first and third arguments, these structure constants are antisymmetric in the upper and lower indices (separately). Since $A_\mu{}_{\al\be}$ and $A_\mu{}^{\al\be}$ on the other hand are symmetric in their three-algebra indices, all terms where they appear vanish. Thus only the first line of (\ref{csterm2}) survives,
\begin{align}
\cL_{\text{CS}}&=\tfrac{1}2\varepsilon^{\mu\nu\rho}
(f^{\al}{}_\be{}^\ga{}_{\de}A_{\mu}{}_{\al}{}^{\be}\partial_{\nu}A_{\rho}{}_{\ga}{}^{\de}
+f^{\al}{}_{\be}{}^{\ga}{}_{\eta} f^{\eta}{}_{\de}{}^\varepsilon{}_\varphi 
A_{\mu}{}_{\al}{}^{\be}A_{\nu}{}_{\ga}{}^{\de} A_{\rho}{}_{\varepsilon}{}^{\varphi})
\end{align}
which is the Chern-Simons term in (\ref{n6chen}).

The potential in (\ref{n5jp}) can be written in many different ways, using the identities
(\ref{bcacablccl})--(\ref{n=5relationer}). To see what happens when we impose the antisymmetry condition on the triple product
it is useful to write it as
\begin{align}
\cL_{\text{pot}}
&=2I_{BAC}-I_{CR}+I_{RR}\nn\\
&=\tfrac23 I_{ABC}-\tfrac13 I_{LL}-\tfrac13I_{RL}+\tfrac13I_{RR}
-\tfrac12(I_{LC}-I_{RC})-\tfrac32(I_{BCA}+I_{ACB}).
\end{align}
The reason for this is that the antisymmetry condition implies the additional identities
\begin{align}
I_{ABC}&=-I_{CBA},&I_{BCA}&=-I_{ACB},&I_{CAB}&=-I_{BAC},\nn
\end{align}
\begin{align}
I_{LC}&=I_{RC},& I_{LL}&=I_{RL}=I_{RR}=I_{LR},
\end{align}
since the left and right of the triple product are treated equally in an $\cN=6$ three-algebra.
Then the $\cN=6$ potential becomes
\begin{align}
\cL_{\text{pot}}&=\tfrac23 I_{ABC}-\tfrac13 I_{RR}\nn\\
&=f^\al{}_\be{}^\ga{}_\eta f^\de{}_\varepsilon{}^\eta{}_\varphi
(-\tfrac23Z^A{}_\al \bar Z_B{}^\be Z^C{}_\ga \bar Z_A{}^\varepsilon Z^B{}_\de \bar Z_C{}^\varphi
+\tfrac13Z^A{}_\al \bar Z_A{}^\be Z^C{}_\ga \bar Z_B{}^\varepsilon Z^B{}_\de \bar Z_C{}^\varphi),
\end{align}
which (after some straightforward massage) agrees with (\ref{n6chen}).

\section{Conclusion}

We have in this paper shown that the $\cN=5$ Lagrangian (\ref{n5chen}) and the $\cN=6$ Lagrangian 
(\ref{n6chen}) can be unified into (\ref{n5jp})
if we instead of an anti-Lie triple system use the corresponding
$\cN=5$ three-algebra. 
The $\cN=6$ Lagrangian then appears as a special case when the triple product is supposed to satisfy an additional antisymmetry condition. We have restricted ourselves to the Lagrangians of the theories, but it should be possible to rewrite
also the supersymmetry and gauge transformations and the equations of motion in the same manner. For the gauge transformations this suggests an exciting possibility in the connection to the associated Lie superalgebra. So far we have only used the Lie superalgebra as a device for keeping track of the contents of the theory -- the gauge and matter fields nicely subsume themselves into the even and odd subspaces, respectively, and thus we can use mathematical results about Lie superalgebras to classify theories of this type.
But maybe the Lie superalgebra associated to the $\cN=5$ three-algebra can somehow also be realized (non-linearly) as a symmetry, like the superconformal algebra of the theory. Then the gauge symmetry would be a part of a bigger symmetry in more or less the same sense as Lorentz invariance is a part of conformal symmetry.
In order to investigate this possibility,
one could try to modify the 
non-linear realization of the (ordinary) Lie algebra associated to a Kantor triple system in \cite{Palmkvist:2005gc}.

\subsubsection*{Acknowledgments}
I would like to thank Jonathan Bagger, George Bruhn, Famin Chen, Andreas Gustavsson, Bengt EW Nilsson, Sung-Soo Kim and Hidehiko Shimada for discussions and comments on the manuscript.
The work is supported by IISN -- Belgium (conventions 4.4511.06 and 4.4514.08), by the Belgian Federal Science Policy Office 
through the Interuniversity Attraction Pole P6/11.



\providecommand{\href}[2]{#2}\begingroup\raggedright\endgroup

\end{document}